\documentclass[%
 superscriptaddress,
 preprint,
bibnotes,
 amsmath,
 amssymb,
 aps,
 prl,
]{revtex4-2}

\usepackage{array}
\usepackage{tabularx}
\usepackage{multirow}
\usepackage{graphicx}
\usepackage{dcolumn}
\usepackage{bm}
\usepackage[colorlinks,
			linkcolor=blue,
            anchorcolor=blue,
            citecolor=blue,
            urlcolor=blue,
            breaklinks=true]{hyperref}

\newcommand{\fref}[1]{Fig.~\ref{#1}}

\newcommand{\p}{\ensuremath{\partial}}

\newcommand{\bra}[1]{\ensuremath{\langle #1|}}
\newcommand{\ket}[1]{\ensuremath{|#1 \rangle}}

\newcommand{\proj}[2]{\ensuremath{\langle #1 | #2 \rangle}}
\newcommand{\tr}{\ensuremath{\mathrm{tr}}}
\newcommand{\calM}{\ensuremath{\mathcal{M}}}
\newcommand{\calL}{\ensuremath{\mathcal{L}}}
\newcommand{\calO}{\ensuremath{\mathcal{O}}}
\newcommand{\ie}{\emph{i.e.}}
\newcommand{\eg}{\emph{e.g.}}
\newcommand{\etc}{\emph{etc.}}

\begin{document}

\title{Quantum-accelerated imaging of N stars}
\author{Fanglin Bao}
\author{Hyunsoo Choi}
\affiliation{Birck Nanotechnology Center, School of Electrical and Computer Engineering, Purdue University, West Lafayette, IN 47907, USA}
\author{Vaneet Aggarwal}
\affiliation{School of Industrial Engineering, and School of Electrical and Computer Engineering, Purdue University, West Lafayette, IN 47907, USA}
\author{Zubin Jacob}\email{zjacob@purdue.edu}
\affiliation{Birck Nanotechnology Center, School of Electrical and Computer Engineering, Purdue University, West Lafayette, IN 47907, USA}
\date{\today}

\begin{abstract}
    Imaging point sources with low angular separation near or below the Rayleigh criterion is important in astronomy, \eg, in the search for habitable exoplanets near stars. However, the measurement time required to resolve stars in the sub-Rayleigh region via traditional direct imaging is usually prohibitive. Here we propose quantum-accelerated imaging (QAI) to significantly reduce the measurement time using an information-theoretic approach. QAI achieves quantum acceleration by adaptively learning optimal measurements from data to maximize Fisher information per detected photon. Our approach can be implemented experimentally by linear-projection instruments followed by a single-photon detector array. We estimate the position, brightness and the number of unknown stars $10\sim100$ times faster than direct imaging with the same aperture. QAI is scalable to large number of incoherent point sources and can find widespread applicability beyond astronomy to high-speed imaging, fluorescence microscopy and efficient optical read-out of qubits.
\end{abstract}

\maketitle
The Rayleigh criterion, which has long hampered the resolution of astronomical imaging, states that two point sources on the image plane with separation smaller than the size of point-spread function (PSF) are irresolvable. However, if stars are modeled by point sources and away from nebulae, textbook signal processing approaches like curve fitting can resolve point sources below the Rayleigh criterion \cite{VANAERT200221,hemmer2016quest}. To do so, a long exposure time is required to suppress the photon shot noise. Nevertheless, long exposure time deteriorates imaging resolution for Earth-based telescopy due to fast-changing atmospheric turbulence \cite{chromey2016measure,fugate1991measurement}. High speed optical imaging that can resolve stars in a short time (\ie, with a few photons) is thus desirable in astronomical imaging, and the key to it is improving Fisher information per detected photon.

Traditional direct imaging uses a photosensor array on the image plane to record intensity. Recent advances in quantum metrology \cite{Tham2017,Yang16,Tsang2016Aug} revealed that direct imaging is inefficient in resolving point sources, in the sense that the extracted classical Fisher information does not saturate the quantum Fisher information bound \cite{Larson2018,PhysRevA.96.062107}. Optimal measurements were reported, against traditional direct imaging, for resolving point sources under certain conditions \cite{Lupo2020Feb,Lu2018December}. However, quantum analyses also shed light on various difficulties in parameter estimation under general conditions, such as, unknown centroid \cite{sajjad2021attaining,Grace2020}, unequal brightness \cite{Prasad2020}, or partial coherence \cite{Liang2021,Larson2018}. Particularly, in the extensive studies which exist, the number of point sources is usually assumed to be known a priori \cite{Yu2018,Zhou2019,Ansari2021} or not scalable \cite{Zhang2020,Lu2018December}. It is an open question to construct a new practical imaging scheme when the entire ground truth is unknown.

Adaptive strategies have proven effective in reconstructing unknown quantum states in quantum state tomography (QST) \cite{Husaer2012,Fischer2000,Hannemann2002}. Adaptive QST seeks to learn optimal measurements from data and uses a few observations to determine the entire state, which is otherwise resource-intensive. Recently, the spirit of iterative re-optimization to approach optimal measurement/control has led to quantum acceleration of frequency estimation \cite{Naghiloo2017}. This is the inspiration behind our fast astronomical imaging in photon-starved conditions. Furthermore, even if a large number of photons are available, quantum acceleration achieved by adaptive modal bases can enable high-speed imaging.

In this study, we propose quantum-accelerated imaging (QAI) for a constellation with unknown number of stars, unknown position, and unknown brightness. Our approach of QAI adaptively learns modal bases from data to maximize Fisher information per detected photon in the sub-Rayleigh region. We demonstrate quantum acceleration factors of around $10\sim 100$ times compared to direct imaging for a given resolution/accuracy. We emphasize that QAI also applies to other incoherent point sources, such as bio-molecules in Brownian motion, blinking quantum dots, and so forth.

We note this advantage is fundamentally different from super-resolution as we are not engineering the PSF. Therefore, we do not term this effect as quantum super-resolution. Far-field super-resolution imaging \cite{Huang2009}, such as, spatially modulated illumination (SMI), confocal laser scanning microscopy (CLSM), stimulated emission depletion (STED), stochastic optical reconstruction microscopy (STORM), \etc, usually involves structured illumination or selective activation/deactivation of fluorescence. In contrast, QAI is fully passive and hence suitable for astronomical imaging. We also show that traditional direct imaging can achieve the same spatial resolution as QAI in the long-measurement-time limit, \ie, with a large number of photons.

\begin{figure*}
    \centering
    \scalebox{1}{\includegraphics{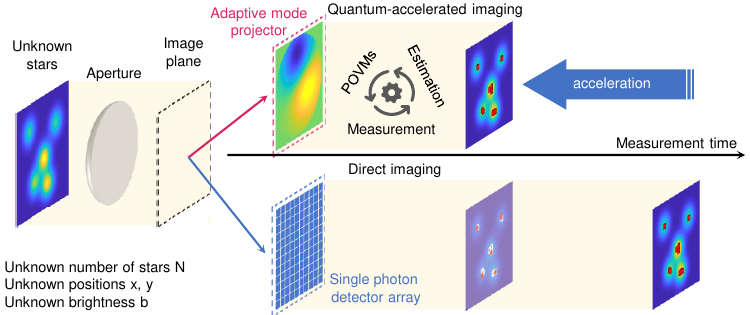}}
    \caption{Schematics of quantum-accelerated imaging (QAI). QAI uses adaptive mode projector on the image plane for projection measurement of the input field, in stark contrast to direct imaging which directly uses single-photon detector array on the image plane to measure the intensity. QAI starts with initial positive operator-valued measures (POVMs), performs measurements, and gets posterior estimations to determine the next optimal POVMs. QAI improves Fisher information per detected photon, and hence accelerates imaging for given resolution down to the sub-Rayleigh region.}
    \label{fig:schematics}
\end{figure*}

\fref{fig:schematics} shows the schematic of our proposed QAI in comparison with direct imaging.
For an unknown constellation scene and fixed aperture, QAI uses an adaptive mode projector as the hardware on the image plane to selectively measure photons in a given modal basis. The adaptive mode projector could consist of a deformable mirror followed by single-photon detectors. The former projects the input optical field onto the given basis while the latter collect excited photons of corresponding modes. The key algorithm of QAI is to find optimal positive operator-valued measures (POVMs) to maximize Fisher information for given amount of photons. Since the ground truth, namely, number of stars ($N$), brightness ($b_n$) and positions ($x_n,y_n$) are unknown, optimal POVMs are approached in an adaptive manner. We start with initial POVMs defined by projections to first 10 Zernike modes. Zernike basis is widely used in image classification, face recognition \cite{app8050827}, and has the same circular symmetry with the aperture. We then approximate the ground truth with posterior estimations to adaptively find subsequent POVMs. This working procedure forms a loop of POVMs-Measurement-Estimation-POVMs, and it only exits when a preset criterion is met, \eg, the fluctuation of consecutive estimations is below a threshold. We note that the size of the POVM set will iteratively update based on prior estimations and might be different with the ground truth number of stars. However, this adaptive approach will eventually converge to the ground truth. This allows us to estimate unknown number of stars, in stark contrast to the literature where number of stars is usually assumed to be known. Remarkably, this adaptive approach consumes less photons, as compared to conventional direct imaging widely deployed in astronomy and bio-imaging.

Now we explain the adaptive algorithm of QAI. As imaging is generally a multi-parameter estimation problem, and the quantum-optimal measurement for multi-parameter estimation remains an open challenge \cite{Lu2021,Gorecki2020optimalprobeserror,Albarelli2019,Liu2019,rehacek2018}, we exploit the single-parameter estimation formalism to propose QAI. In the weak-signal regime where the detector only receives, at most, one single photon at one time, one-photon state of the optical field for $N$ incoherent point sources is described by the following mixed-state density operator
\begin{equation}
    \rho = \sum_{n=1}^N b_n\ket{\phi_n}\bra{\phi_n},
\end{equation}
where $b_n$ is the brightness of the $n_\mathrm{th}$ star with normalization condition $\sum_nb_n=1$. The mean image
\begin{equation}
    |\proj{x,y}{\phi_n}|^2 = \Psi(x-x_n,y-y_n,\sigma)
\end{equation}
of the $n_\mathrm{th}$ star on the image plane is given by PSF $\Psi$ of width $\sigma$ located at $(x_n,y_n)$. Note that $\proj{\phi_n}{\phi_n}=1$ but $\proj{\phi_m}{\phi_n}\neq 0$ for $m\neq n$, due to the wavefunction overlap when separation of stars is below the Rayleigh criterion. We use singular value decomposition (SVD) to find the orthonormal bases $\{ \ket{O_n} \}$ for $\{ \ket{\phi_n} \}$, $\proj{O_m}{O_n}=\delta_{mn}$, where the density operator has an $N\times N$ matrix form. We further diagonalize the density matrix $\rho$ to get its eigenbases $\{ \ket{e_n} \}$, $\proj{e_m}{e_n}=\delta_{mn}$, so that
\begin{equation}
    \rho = \sum_{n=1}^N D_n \ket{e_n}\bra{e_n},
\end{equation}
where $D_n$ is the corresponding eigenvalues.

For a parameter $\mu$ to be estimated, $\mu\in \{ x_n,y_n,b_n \}$, the symmetric logarithmic derivative (SLD) operator $\calL_\mu$ is defined by the following equation,
\begin{equation}
    2\p_\mu \rho = \calL_\mu \rho + \rho \calL_\mu,
\end{equation}
and its solution in the eigenbases of $\rho$ is
\begin{equation}
    \calL_\mu = \sum_{mn}\frac{2}{D_m+D_n}\bra{e_m}\p_\mu\rho \ket{e_n}\cdot \ket{e_m}\bra{e_n}.
\end{equation}
The single-parameter quantum Fisher information about $\mu$, $K_{\mu\mu}$, that one can retrieve from one photon is
\begin{equation}
    K_{\mu\mu} = \tr[\rho \calL_\mu \calL_\mu].
\end{equation}
Denote the parameter of maximum quantum Fisher information as $\nu \equiv \mathrm{argmax}_\mu[K_{\mu\mu}]$. The eigenbases of $\calL_\nu$, $\{ \ket{f_n} \}$, gives the near-optimal bases for estimating parameter $\nu$. Therefore, we propose the following POVMs
\begin{equation}
    \calM_n = \ket{f_n}\bra{f_n},\quad n=1,\cdots,N
\end{equation}
in this letter to approach the optimal measurements.

\begin{figure}
    \centering
    \scalebox{1}{\includegraphics{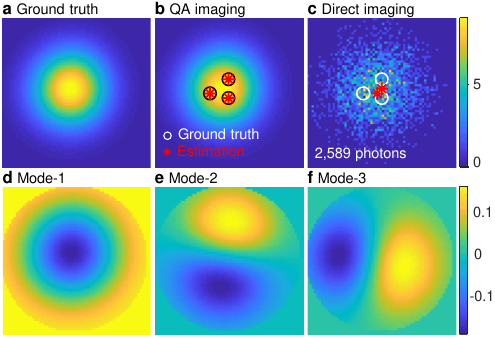}}
    \caption{Demonstration of QAI for given photons showing accuracy enhancement against direct imaging. a, Ground truth scene of 3 stars with unequal brightness. b, Quantum-accelerated imaging of stars (red star) with 2,589 photons is consistent with the ground truth (circles). c, Direct imaging of stars with 2,589 photons shows error in number of stars and inaccurate positions. Conventional Gaussian mixture model is used in direct imaging. d-f, Sample profiles of adaptive modes used in QAI that improve Fisher information.}
    \label{fig:3star}
\end{figure}

\fref{fig:3star} shows the demonstration of our proposed QAI for 3 stars with unequal brightness gathering together within one cluster. The maximum distance among 3 stars is below one $\sigma$ in the sub-Rayleigh region, and the unknown parameter set is $\Theta=\{N,x_n,y_n,b_n | n=1,\cdots,N\}$. \fref{fig:3star}b shows that with 2589 photons QAI is able to image the scene comparable to the ground truth (\fref{fig:3star}a). In contrast, conventional direct imaging (\fref{fig:3star}c) with the same amount of photons predicts wrong number of stars (2 vs. ground truth 3) and inaccurate star positions. For better estimation accuracy, here we implement the posterior estimation with the Bayesian inference for both QAI and direct imaging. For a given prior probability mass function (PMF), $p_k(\Theta)$, in the parameter space and observed new data $z$, the posterior PMF is given by the Bayesian update
\begin{equation}
    p_{k+1}(\Theta|z) = \frac{p(z|\Theta)p_k(\Theta)}{\sum_{\Theta'} p(z|\Theta')p_k(\Theta')},
\end{equation}
where $k$ indicates the $k_\mathrm{th}$ iteration. Maximum likelihood estimation is then applied to $p_{k+1}(\Theta)$ to get the posterior estimation and find adaptive POVMs. \fref{fig:3star}d-f show adaptive mode profiles in the last iteration of QAI, given by $\proj{x,y}{f_n}$. For a full illustration of QAI, see Visualization 1 in supplementary material for adaptive processes. In comparison with direct imaging which accumulates photons pixel by pixel, mode projection like \fref{fig:3star}d-f improves Fisher information per detected photon as explained above. In Monte Carlo simulations of measurement, we use the Poisson distribution to generate random number of photons around the mean count. For QAI, the mean count is given by $\tr[\rho \calM_n]$. For direct imaging, the mean count is given by $\bra{x,y}\rho \ket{x,y}$.

Without loss of generality, we have discretized the image plane into a $64\times64$ pixel/mirror array and defined the maximum inner tangent circle as the unit disk of the imaging space. The width of PSF is $\sigma=10$ pixels. For the purpose of reducing computational complexity of Bayesian update, we assume that the maximum number of stars is 3 and the maximum radial position is $\sigma$. Also, we use 5 discrete radial grids, 4 angular grids, and 5 brightness levels to reduce the dimensionality and set the spatial/brightness resolution.

\begin{figure}
    \centering
    \scalebox{1}{\includegraphics{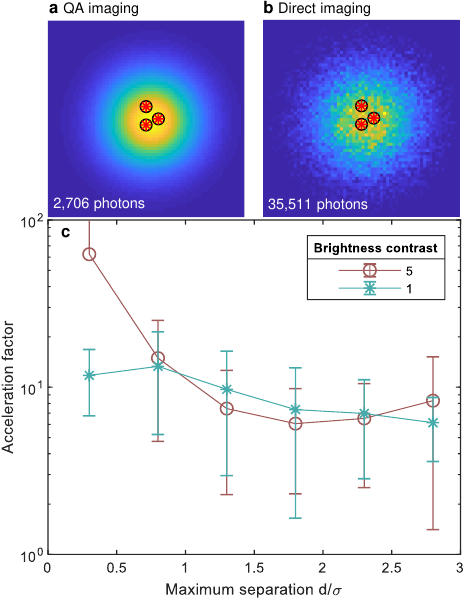}}
    \caption{Demonstration of QAI for preset accuracy showing acceleration against direct imaging. a, QAI of 3 stars (red star) consumes 2,706 photons to hit the ground truth (black circle). b, Direct imaging of the same 3 stars consumes 35,511 photons to hit the ground truth. QAI shows an acceleration over 10 folds in measurement time for fixed aperture. c, Acceleration factor for 12 different scenes shows $10\sim 100$ times universal acceleration of QAI. $d$ is the maximum distance among 3 stars. $\sigma$ is the width of the point-spread function. Brightness contrast is the ratio of the highest brightness over the lowest brightness among 3 stars. 12 runs of imaging have been performed for each scene for statistics.}
    \label{fig:statistics}
\end{figure}

Now we show the overall quantum acceleration factor of our proposed QAI against direct imaging. For given estimation resolution/accuracy, the acceleration factor is defined as the ratio of photons consumed by direct imaging over photons consumed by QAI. \fref{fig:statistics}a and b are the final imaging results of QAI and direct imaging, respectively, when the estimation hits the ground truth. QAI consumes 2,706 photons, in contrast to 35,511 photons used in direct imaging, to correctly resolve 3 stars in the sub-Rayleigh region. The overall 1 order in magnitude of acceleration holds for various scenes. \fref{fig:statistics}c shows the statistics of the acceleration factor for 12 different scenes (6 different position configurations times 2 different brightness configurations). Each scene is tested with 12 independent runs, and the mean (curves) and standard deviation (error bars) of acceleration factor are derived from these 12 values. Due to the shot noise, the acceleration factor fluctuates. We emphasize that the acceleration effect is robust as the acceleration factor is significantly over 1. Moreover, the acceleration factor shows an increasing trend with respect to small separation and large brightness contrast, making QAI especially promising for resolving point source clusters, \eg, searching for exoplanets and identifying single-photon nano-emitters. Finally, we point out that the quantum acceleration effect should not be confused with conventional super-resolution. Conventional super-resolution imaging, especially those based on active PSF engineering, is applicable to general images, while QAI uses the prior of point sources. With the point-source prior, direct imaging can also achieve sub-Rayleigh resolution in the long-time measurement limit, as shown in \fref{fig:statistics}b.

\begin{figure}
    \centering
    \scalebox{1}{\includegraphics{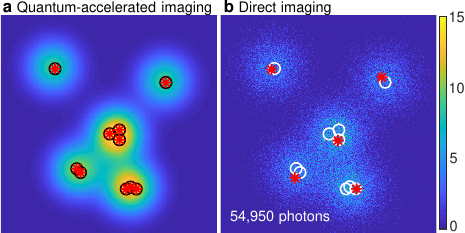}}
    \caption{QAI scaled up for 10 stars. With 54,950 photons QAI shows improved accuracy against direct imaging. Especially QAI correctly estimates number of stars, even in sub-Rayleigh clusters with limited photon budgets. Circle: ground truth positions. Red star: estimated position. Direct imaging is based on conventional CLEAN algorithm.}
    \label{fig:scalingup}
\end{figure}

QAI can be scaled up to large number of point sources. The computational complexity of SVD scales as $\calO(mN^2)$, where $m$ is the number of pixels of the imaging space assuming $m>N$. The eigen-decomposition scales as $\calO(N^3)$. Therefore, the overall complexity for searching POVMs is $\calO(mN^2)$. The Bayesian update is computationally expensive and scales exponentially as $\calO((IRA)^N)$. Here, $I, R, A$ are number of discrete brightness, radial and angular grids. Within 3 stars in this letter, a standard PC can complete QAI within about 10 seconds. To scale up to 10 or more stars, the Bayesian update for posterior estimation must be replaced with more efficient gradient descent approaches. Here, we provide an alternative scaling approach that splits the scene into star clusters. To do so, we utilize the conventional CLEAN algorithm \cite{hogbom1974aperture} as pre-processing. We start with direct imaging, 1) search for the brightest pixel in the scene, 2) subtract a PSF multiplied by loop gain, 3) repeat 1)-2) until the sidelobe of the peak pixel decreases to desired threshold, and 4) assemble PSFs into exhaustive and exclusive clusters. Consequently, the aforementioned QAI can be applied to each cluster where number of stars is usually small and below 3. This scaling approach is linearly proportional to the number of stars. \fref{fig:scalingup}a shows QAI scaled up for 10 stars based on 10,000-photon CLEAN pre-processing and 54,950 photons in total. In comparison, \fref{fig:scalingup}b shows the CLEAN estimation with a total budget of 54,950 photons. CLEAN is poor at resolving stars in the sub-Rayleigh region, but efficient in locating blur spots/clusters, suitable as the pre-processing of QAI for scaling up.

We have proposed the quantum-accelerated imaging for unknown stars. We have demonstrated around $10\sim 100$ times quantum acceleration in measurement time against traditional direct imaging, by adaptively searching optimal measurements to maximize Fisher information per detected photon. Our quantum-accelerated imaging is robust and also applicable to other incoherent point sources with unknown number, unknown positions, and unknown brightness. Our results call for future research in fast optical imaging that can adapt quantum acceleration to general extended objects beyond point sources.

\bibliography{qai}

\end{document}